\documentclass[journal]{vgtc}                     

\onlineid{0}

\vgtccategory{Research}

\vgtcpapertype{please specify}

\title{Task Breakpoint Generation using Origin-Centric Graph\\
in Virtual Reality Recordings for Adaptive Playback}

\author{%
  \authororcid{Selin Choi}{0009-0007-1370-5359},
 \authororcid{Dooyoung Kim}{0000-0002-6003-2181},
  \authororcid{Taewook Ha}{0009-0004-7946-8577},
  \authororcid{Seonji Kim}{0000-0003-3153-6177}, and
  \authororcid{Woontack Woo}{0000-0002-5501-4421}
  }

\authorfooter{
  \item
  	Selin Choi is with KAIST UVR Lab.
  	E-mail: selin.choi@kaist.ac.kr
  \item
  	Dooyoung Kim is with KAIST KI-ITC ARRC.
  	E-mail: dooyoung.kim@kaist.ac.kr
    \item
  	Taewook Ha is with KAIST UVR Lab.
  	E-mail: tw.ha@kaist.ac.kr
    \item
  	Seonji Kim is with KAIST UVR Lab.
  	E-mail: seonji.kim@kaist.ac.kr

  \item Woontack Woo is with KAIST UVR Lab, KAIST KI-ITC ARRC.
  	E-mail: wwoo@kaist.ac.kr
}

\abstract{
We propose a method for generating task breakpoints based on an \textit{Origin-Centric Graph (OCG)} to segment goal-oriented activity recordings into task units for adaptive playback in Virtual Reality (VR) environments. With the development of Augmented Reality (AR)/VR head-mounted displays (HMDs), research on adaptive tutorials and authoring tools has become active, but existing task segmentation methods mainly rely on manual annotation or are restricted to 2D video which limits their applicability to 3D VR contexts. In our approach, assembly scenarios with clearly defined task boundaries are recorded using a structured spatio-temporal scene graph (STSG), and the OCG is employed to track changes in the central object and the formation of new groups, thereby generating task breakpoints automatically. A user study collected user-perceived task breakpoints to establish ground truth (GT), and comparison with the algorithm-detected breakpoints demonstrated high agreement and confirmed accuracy in supporting adaptive playback. The proposed task segmentation method provides a foundation for dynamically adjusting VR playback according to user proficiency and progress, with potential for extension into automatic timeline segmentation systems for diverse VR recordings.
}


\teaser{
  \centering
  \includegraphics[width=\linewidth]{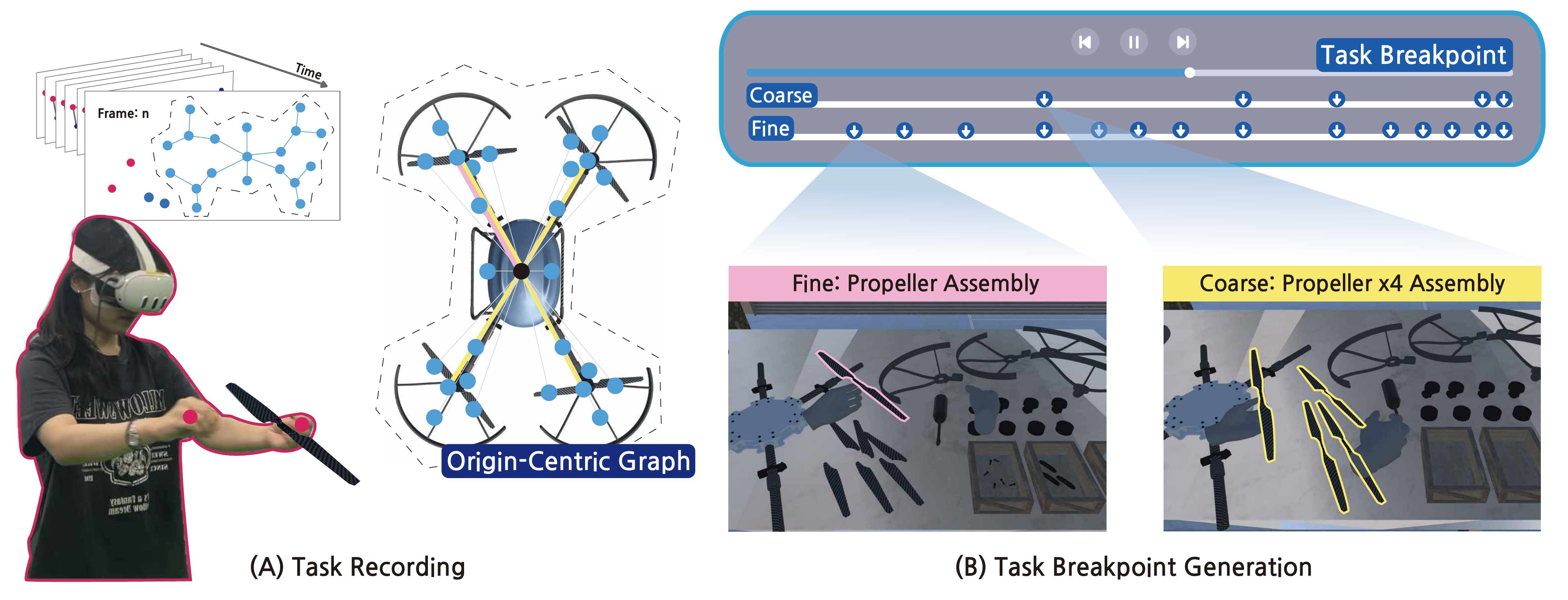}
  \caption{%
  	The figure illustrates how tasks are recorded in VR and segmented into fine and coarse units through task breakpoint generation. (A) shows the recording of a drone assembly process using the STSG and the construction of the OCG. (B) presents the result of task breakpoint generation: the pink marker indicates a fine breakpoint, such as the completion of assembling a single propeller, while the yellow marker indicates a coarse breakpoint, such as the completion of assembling all four propellers as a single unit.%
  }
  \label{fig:teaser}
}

\graphicspath{{figs/}{figures/}{pictures/}{images/}{./}} 

\usepackage{tabu}                      
\usepackage{booktabs}                  
\usepackage{lipsum}                    
\usepackage{mwe}                       
\usepackage{amsmath, amssymb}
\usepackage{caption}
\usepackage{multirow}
\usepackage{mathptmx}                  
\usepackage{enumitem}

\begin{document}


\firstsection{Introduction}

\maketitle

As spatial computing technology emerges as the next-generation media, the importance of spatial video, which records and replays experiences in 3D beyond traditional 2D screen viewing, is increasing. Spatial video recorded in Virtual Reality (VR) environments allows users to explore scenes in three dimensions from various angles and distances, unlike 2D video constrained to fixed viewpoints. Moreover, it provides immersive 1:1 scale experiences as if users are physically present, making it applicable in a wide range of learning contexts from everyday tasks to professional activities~\cite{cooper2021transfer, horvath2021analysis}. With advances in head-mounted displays (HMDs) enabling precise tracking of user actions, adaptive playback tailored to individual proficiency and performance speed has become feasible~\cite{huang2021adaptutar}. Supporting such adaptive learning requires segmenting spatial video into meaningful task units hierarchically structured according to how users perceive tasks, similar to chapter-based segmentation in online video. However, existing task recording and tutorial approaches require manual segmentation, demanding excessive time and effort, and most automatic segmentation methods are limited to 2D video and thus hard to apply to 3D spatial video. Therefore, to effectively utilize spatial video, it is necessary to structurally record spatio-temporal information of task processes and develop a hierarchical segmentation method that can adaptively respond to learner performance.

Recently, tutorial systems that provide guidance or feedback through spatial video recording using VR/Augmented Reality (AR) HMDs have been actively studied\cite{liu2023instrumentar, stanescu2023state, stanescu2024error, thanyadit2022xr}. In particular, adaptive playback features that enable section repetition or speed adjustment according to the user's skill level or task progress are being vigorously developed, mainly focusing on VR/AR-based systems that respond in real time to task execution~\cite{zhang2025integrative, hajahmadi2024investigating,  kim2019evaluating, maloney2013investigating, gutierrez2011adaptive, bimba2017adaptive, huang2021adaptutar}. However, most of these systems require manual segmentation of task units or stepwise recording, which incurs high cost and time. While humans naturally and effectively learn complex tasks by dividing them into fine and coarse units~\cite{newtson1973attribution,newtson1976perceptual}, existing video segmentation studies~\cite{carreira2017quo, feichtenhofer2019slowfast} and automatic tutorial generation studies~\cite{petersen2012learning, petersen2013real, young2021automated} segment tasks only at the smallest action unit, limiting meaningful fine/coarse segmentation. Additionally, most existing automatic segmentation methods are designed based on RGB and depth images~\cite{petersen2012learning, petersen2013real}, making it difficult to effectively extract meaningful task units in VR environments where user's viewpoints can freely change and interactions occur in real time. To enable automatic task segmentation in VR, it is essential to adopt a new approach that records spatial, object, user, and interaction data in a structured manner and applies hierarchical analysis to extract task units.

We propose a method that applies a scene graph-based VR recording approach to capture spatial videos of goal-oriented tasks with adaptive playback capabilities, and utilizes an \textit{Origin-Centric Graph (OCG)} to automatically generate task breakpoints. A task breakpoint is defined as a temporal boundary that segments a task interval into a meaningful unit.
This study focuses on assembly scenarios with relatively clear task distinctions. Through a preliminary study, criteria and decisive information for users’ recognition of task units were collected, and it was shown that adaptive playback of assembly tasks requires segmentation not only into single processing units but also into fine and coarse units. To effectively represent user-object interactions over time, a data structure was designed based on a spatio-temporal scene graph (STSG) that uses nodes containing six degrees of freedom (6DoF) information of objects and hands, enabling precise expression of user actions and task status. Furthermore, an OCG was proposed to capture assembly structures and hierarchical changes, and a technique was developed to automatically generate task breakpoints by comparing OCG with STSG, thereby segmenting tasks into fine and coarse units. Finally, a post-processing step reflecting user action completion was added to refine the automatically generated task units to better align with user-perceived task intervals.

We conducted an evaluation to verify the effectiveness of the proposed task segmentation algorithm based on task breakpoint generation. Two assembly task videos with different levels of complexity (bicycle and drone), recorded by VR experts, were used. We compared user experiment data ($n=24$) where task breakpoints were identified as ground truth (GT) against the breakpoints automatically generated by our algorithm. Through DBSCAN clustering, fine and coarse GT breakpoints were determined from participants’ responses, and using the time ranges reported by participants as criteria, we calculated the detection accuracy and timing errors of our proposed task segment detecting algorithm. The detection of fine breakpoints showed high accuracy, with an overall F1 score of 0.98, while coarse breakpoints demonstrated stable performance with an overall F1 score of 0.90. These results indicate that the proposed algorithm detects task breakpoints consistent with user perception with high accuracy, demonstrating reliable segmentation performance for adaptive playback. Additionally, post-experiment interviews confirmed that participants could experience the recorded VR tasks immersively. Taken together, these findings suggest that meaningful task structure can be reliably inferred directly from VR interaction recordings, establishing a practical foundation for adaptive playback without requiring manual annotation. This evaluation confirms that our VR recording approach can effectively generate breakpoints necessary for adaptive playback through an initial task segmentation process, enabling immersive VR task recording experiences.

The contributions of this work are threefold. First, we propose a STSG-based VR recording method that enables recording of spatial videos for adaptive playback. Our method can automatically record VR videos segmented into task units generating task breakpoints as long as the VR content provides basic object and interaction information, without requiring additional sensors.
Second, we propose a data structure that effectively represents hierarchical information required for adaptive playback by utilizing the STSG to capture user-object interactions and the OCG to capture assembly structure changes.
Finally, we present an algorithm that automatically generates breakpoints to segment task units necessary for adaptive playback, and we demonstrate through user experiments that it accurately segments task units.
By using the proposed VR recording system, spatial videos can be automatically and hierarchically segmented into tasks without manual effort, significantly reducing the time and labor required for creating goal-oriented learning content. Specifically, it reflects hierarchical structures as perceived by general users, capturing both fine and coarse units, enabling effective use for adaptive playback.
Moreover, since the system uses information already available in existing VR content, it can be extended beyond tutorial videos to various domains and contribute to cost-effective generation of spatial videos with adaptive playback capabilities.

\section{Background}

\subsection{VR/AR Tutorial}
Existing tutorial systems for VR and AR environments primarily focus on providing instructions to users or supporting learning through feedback on task outcomes~\cite{stanescu2023state, stanescu2024error, zhang2025integrative, hajahmadi2024investigating}. However, in domains where user actions and postures are critical or task structures are complex, it is essential to observe the sequence of actions continuously over time~\cite{zikas2023mages, eckhoff2018tutar, thoravi2019tutorivr}. For example, in tasks that involve handling small components or tool operations, pre-recorded, demonstration-based tutorials are effective, and related research has been actively conducted~\cite{thoravi2019tutorivr, chidambaram2021processar, chidambaram2022editar}. Recording-based tutorials allow experts to capture demonstrations and replay them spatially in VR so that users can follow along~\cite{cho2023time, cho2023hybrid, thanyadit2022xr}. They also provide the benefit of repeated playback, enabling learners to review specific segments as needed. Such recording functionality in AR/VR can continuously capture user movements and environmental context in three dimensions, and replay them from multiple perspectives, thereby offering a more intuitive and effective learning experience than traditional instructor-based tutorials~\cite{cooper2021transfer, horvath2021analysis}.

Nevertheless, recording-based tutorials pose the drawback that the learner’s pace may not match the playback speed, forcing users to repeatedly search for relevant segments. This issue becomes even more pronounced in VR spatial video, which contains diverse 3D information. Consequently, there is a growing need for adaptive replay techniques that support repeated segment review and playback control according to learners’ progress and performance~\cite{kim2019evaluating, maloney2013investigating, gutierrez2011adaptive, bimba2017adaptive}. Recent studies have explored methods such as automatically pausing playback according to task progress~\cite{thanyadit2022xr} or adaptively adjusting tutoring elements~\cite{huang2021adaptutar}. However, these approaches typically require experts to manually record each task step or to segment demonstrations into meaningful units, leaving task segmentation a core challenge that still demands significant manual effort.

\subsection{Task Segmentation for Adaptive Playback}
In adaptive playback systems, temporal segmentation plays a crucial role when considering factors such as task flow and learning pace. Segmented steps help learners more easily understand and effectively master complex procedures. In the field of computer vision, extensive research has been conducted on action recognition and segmentation based on 2D video~\cite{carreira2017quo, feichtenhofer2019slowfast}. However, most approaches focus on individual action units, insufficiently capturing the relationships between actions. In learning scenarios such as recording-based tutorials, where the contextual continuity between steps is essential, modeling such relationships becomes critical. To address this limitation, recent approaches have employed graph-based structures to model inter-action relationships for more systematic analysis~\cite{ji2020action, bar2020compositional, luo2021moma}. In AR, methods have been proposed to segment video sequences or live streams into task units to generate and update instructional documents~\cite{petersen2012learning, petersen2013real, young2021automated}. Nonetheless, these approaches remain largely limited to 2D video, failing to capture the diversity of user perspectives and motion, and focusing primarily on single action units rather than effectively segmenting the overall task structure. In short, while automated segmentation techniques have advanced steadily in 2D video, efforts that leverage 3D spatial data are still scarce.

As discussed above, prior research has largely remained focused on segmenting individual action units, which does not adequately reflect the complete structure of tasks. Yet learning processes are inherently complex and hierarchical, making simple unit-level segmentation insufficient. Research in cognitive science and psychology demonstrates that people naturally organize goal-directed behavior hierarchically, distinguishing between fine and coarse units and recognizing the significance of transition points~\cite{newtson1973attribution,newtson1976perceptual}. Building on these insights, many studies have proposed segmenting tasks into meaningful fine and coarse units at critical transition points, applying such segmentation to automatic instruction generation and to improving the efficiency of human–computer interaction~\cite{mura2013ibes}. Moreover, AR research has visualized causal relationships between higher-level event units and fine-grained interaction units, showing that each unit plays a distinct role and exerts different effects on learning~\cite{jain2025visualizing}. Thus, structuring goal-oriented tasks hierarchically provides a foundation for recording not only sequential steps but also higher-level goals and their constituent substeps.

\subsection{Spatial Video Recording}
Previous research on VR recording has focused on storing behavioral data, gaze, and user input to support experiment replication and behavioral analysis~\cite{javerliat2024plume, hubenschmid2022relive}. However, such approaches have not sufficiently captured fine-level details such as task execution processes or object state changes. For example, some work has recorded spatial design processes to allow users to navigate timelines~\cite{mahadevan2023tesseract}, but these efforts concentrated only on spatial aspects and could not comprehensively represent user actions. Other studies have proposed authoring environments that simplify the creation of AR, VR, or video content through single-shot demonstrations~\cite{chidambaram2022editar}, or methods that automatically generate AR tutorials from recorded data~\cite{liu2023instrumentar}. Yet these approaches rely heavily on physical sensors for task segmentation, resulting in low reproducibility. Therefore, a more comprehensive recording method is required—one that can capture all information necessary for both user actions and task execution, supported by a structured storage format that clearly reflects task composition.

VR spatial video, when analyzed solely through image-based methods, cannot capture user–object interactions or changes in object-to-object relationships. This limitation highlights the need for a representation that structurally encodes user actions and task states within 3D data. Spatio-temporal scene graphs provide such a framework, representing spatial data over time by modeling the relationships between objects and environments, which is advantageous for tracking task progress~\cite{ji2020action, qiu2023virtualhome}. Nonetheless, prior work has primarily focused on the spatial relationships of large objects, leaving fine-level object-to-object relations in scenarios such as assembly underrepresented~\cite{tahara2020retargetable, kim2024object, kim2024spatial}. Although state graphs have been proposed to capture assembly states~\cite{stanescu2023state, stanescu2024error}, they primarily address object states and do not incorporate continuous user actions or interactions. Similarly, research has modeled user–environment–object interactions in 3D environments~\cite{li2023generating}, but the focus on generating short action snippets limits their utility as comprehensive task recording methods. Consequently, to effectively record tasks and hierarchically segment them into fine and coarse units, a holistic recording approach is required—one that integrates user actions, object properties, and task structures.

\section{Methodology}
In this study, we propose a method to automatically generate a task-segmented VR recording by applying an OCG-based breakpoint generation to the STSG of a VR expert's task performance. We also proposed a VR recording method that extracts user hand movements, object state changes, and object-to-object interactions at each frame, and stores them in an integrated scene graph representation. Based on this representation, an algorithm segments the task into fine and coarse units. The scene graph accurately records relationships such as user grasp/release of objects, object connections, and tool operations. The recorded data are then processed by the algorithm to identify fine and coarse breakpoints, which are aligned with the completion of user actions. This ensures that frame-level object states and interaction changes are systematically and structurally managed. We selected assembly tasks as representative scenarios for task-segmented spatial video recording, as they clearly reveal structural state changes and allow for easy assessment of task progress. The following sections first present a preliminary study that organizes user-defined criteria for task segmentation, and then describe the scene graph data structure for VR recording along with the OCG–based algorithm for task breakpoint generation.

\subsection{Preliminary Study}
\subsubsection{Study Design and Procedures}

To design a recording and segmentation method for adaptive replay, we conducted a preliminary study aimed at structuring spatial video data and identifying criteria for meaningful video segmentation from the user’s perspective. The experiment used the assembly of an IKEA\footnote{https://ikea.com} chair as the representative scenario. The study was reviewed and approved by the KAIST Institutional Review Board (IRB; Protocol No. KH2024-180).
We recruited four expert participants (age: $M=28.5$, $SD=1.73$; two females, two males) with extensive VR research experience and familiarity with VR interactions to record the assembly process in a VR environment. The assembly task was designed, with reference to the original instruction manual and prior work~\cite{funk2015benchmark}, to support both sequential and parallel assembly orders. Participants wore a Meta Quest~3 VR HMD\footnote{https://www.meta.com/quest/quest-3/}, and using a Unity-based recording system, their hand and object movements were captured in 6DoF at 60 frames per second. A one-hour training session was provided beforehand to ensure sufficient familiarity with the assembly procedure. The chair assembly videos (sec: $M=123$, $SD=9.2$) recorded by the four experts were then utilized as spatial videos to derive task segmentation criteria and the information necessary for adaptive replay, under the assumption that general users would follow these recordings during task execution.

In this study, a task breakpoint is defined as a moment of cognitive transition during task execution, serving as the starting point of meaningful segmentation units for adaptive replay. To investigate such breakpoints, we conducted a preliminary study with 12 general users (age: $M=26.0$, $SD=3.74$; six females, six males) from diverse backgrounds, recruited from local undergraduate and graduate students through the official KAIST community website. Participants, whose VR experience ranged from complete novices to moderately experienced users, viewed four VR assembly recordings produced by experts and performed segmentation tasks by dividing the workflow into fine and coarse units.
Participants were instructed to indicate subjectively meaningful breakpoints in the continuous recordings, marking them at two different levels. A coarse breakpoint was defined as a higher-level unit encompassing subordinate steps, typically associated with overall context or goal changes, while a fine breakpoint was defined as a lower-level unit reflecting specific detailed actions or subgoal transitions~\cite{kurby2008segmentation, hard2006making}. Participants wore a Meta Quest~3 headset and first watched each VR video in its entirety to understand the overall task structure. They then rewatched the same video to identify fine breakpoints, followed by coarse breakpoints. Each participant evaluated all four videos, and the order of presentation was counterbalanced using a Latin square design. After completing the segmentation tasks, participants were interviewed about their segmentation process and criteria. The entire study lasted approximately 1.5 hours, and each participant received a \$10 reward. Based on the collected fine and coarse breakpoint data and interview results, we extracted the key characteristics of breakpoint identification.

\subsubsection{Results and Implications}

\textbf{Task Segmentation.} We analyzed the collected fine and coarse breakpoint annotations along with interview responses to extract key characteristics of breakpoint recognition. Many participants commonly identified task completion moments, particularly the completion of assembly actions, as meaningful breakpoints. For fine breakpoints, participants often perceived transitions relative to a central object rather than every simple connection.
To bridge the gap between human perception and algorithmic implementation, we abstracted these observations into three distinct structural triggers, which form the basis of our computational heuristics:
\begin{enumerate}[label=\textbf{T\arabic*.}, leftmargin=3em]
    \item \textbf{Integration-based Transitions:} Breakpoints were frequently marked when a part was connected to the central object with the highest number of connections in the final assembly. This indicates that users perceive the integration of a component into the structural core as a significant milestone.
    
    \item \textbf{Centrality-based Transitions:} Even when a part was connected sequentially to the previous a group, participants recognized a breakpoint if the connection was closer to the central object. This suggests that the relative hierarchical importance of a connection influences task segmentation.
    
    \item \textbf{Topology-based Transitions:} When a part was connected to a new group rather than the existing one, the preceding step was identified as a breakpoint. Users viewed the formation of a new sub-assembly as a distinct sub-task.
\end{enumerate}

Interview results reinforced these findings, highlighting that new group formation and connections to central objects were regarded as important criteria ($n=7$ out of 12). 
Coarse breakpoints, in contrast, were perceived as aggregations of multiple fine units ($n=10$). Repetitive fine actions that could proceed in parallel were often grouped into one coarse unit, and participants ($n=8$) emphasized shifts in task goals or overall task context as key segmentation criteria.
Both fine and coarse breakpoints were typically aligned with the completion of an action, with tool-based operations (e.g., tightening, fastening) particularly likely to mark breakpoint moments. This finding provides the logical basis for our behavior-based refinement process, which adjusts the algorithmic timestamps to match the user's cognitive release point rather than the exact moment of physical contact.

\noindent\textbf{Data Structure.} 
To develop a task-segmented recording method, it is necessary to move beyond simple time-based recording and instead employ structured representations that explicitly capture task relationships. Assembly tasks involve continuous state changes of objects and complex tool-based interactions over time, making frame-level STSGs especially suitable for capturing task progress. The preliminary study revealed that participants consistently marked breakpoints at the completion of major assembly actions or tool operations, underscoring the importance of continuously recording user-–object interaction events such as grasp/release and tool manipulations. To effectively represent sequential workflows and breakpoint transitions, a graph-based data structure is required that encodes user actions and interactions relative to central objects while capturing structural transitions around them. Building on these insights, we propose an STSG-based recording approach that represents object state changes and user–-object interactions over time, enabling explicit identification of breakpoint units to support adaptive playback through a structured representation of task workflows.

\subsection{Spatio-Temporal Scene Graph for Task Recording}
Based on the findings of the preliminary study, we propose an STSG for data recording. The STSG is defined as $G_t = (V, E_t)$ at each frame $t$, where $V$ represents the set of nodes and $E_t$ denotes the set of edges. Unlike conventional scene graphs that primarily focus on static spatial proximity, our edge set $E_t$ encodes both the structural connectivity between objects and the functional interactions between the user's hands and those objects. By capturing these distinct relationships within a single unified structure, the STSG records not only the relationships among objects but also user–object interactions, thereby providing a quantitative basis for detecting meaningful task breakpoints and action completions.

The node set $V$ is defined as the union of user nodes $V_U$ and object nodes $V_O$ (i.e., $V = V_U \cup V_O$), which are pre-assigned before the recording begins. 
The user nodes $V_U = \{u_{left}, u_{right}\}$ represent the user's hands, where each node stores a set of 6DoF poses for 21 hand joints, denoted as  $\{J_{i,t}\}_{i=1}^{21}$. Here, each $J_{i,t}$ represents the position and rotation of the $i$-th joint provided by the hand tracking feature of the Meta Quest~3 SDK.
The object nodes $V_O = \{o_1, o_2, \dots, o_N\}$ represent $N$ components and tools. Each node $o_i \in V_O$ contains static attributes (ID, name, and $Category \in \{part, tool\}$)  and a dynamic 6DoF pose updated at 60 frames per second, while the static attributes remain fixed after the initial assignment.

The edges $E_t$ are defined by the Hand Adjacency Matrix, which represents the relationships between the user and objects, and the Adjacency Matrix, which represents the relationships between objects. Each element in these matrices corresponds to the weight of an edge in the graph and is recorded at every frame. Specifically, the user-object relationship is represented by the Hand Adjacency Matrix $H_t \in \mathbb{R}^{2 \times N}$:
\begin{equation}
H_{ij,t} = 
\begin{cases} 
1 & \text{if hand } i \text{ is grasping object } j \text{ at frame } t, \\ 
0 & \text{otherwise,} 
\end{cases}
\label{eq:HandMatrix}
\end{equation}
where the two rows represent the left and right hands, respectively, and \(N\) is the total number of object nodes. This matrix is designed to accurately capture the moment when a user grasps an object by recording a value of 1 for the corresponding element when the user's hand contacts an object, and 0 otherwise.
Simultaneously, the relationships among the $N$ components are represented by the Adjacency Matrix $A_t \in \mathbb{R}^{N \times N}$:
\begin{equation}
A_{ij,t} = 
\begin{cases} 
1 &
\begin{array}{l}
\text{if part } i \text{ and part } j \text{ are connected,} \\
\text{or if tool } i \text{ is manipulating part } j,
\end{array} \\
0 & \text{otherwise,} 
\end{cases}
\label{eq:AdjMatrix}
\end{equation}
\cref{eq:AdjMatrix}, this \(N \times N\) square matrix represents the connections among all \(N\) components (both parts and tools) within the VR recording. Each element indicates whether a physical connection exists between two components; when parts are coupled or when a tool is in the process of manipulating a part, the corresponding element is marked as 1, and 0 otherwise.

By recording the adjacency matrices defined in \cref{eq:HandMatrix} and \cref{eq:AdjMatrix} on a per-frame basis, the system can comprehensively encapsulate the connectivity between parts as well as tool-mediated manipulations occurring during the assembly task. Consequently, this approach allows for the systematic and structural recording of various relationships, such as user-object grasps, inter-part connections, and tool operations, while providing a quantitative basis for the explicit analysis of user behavior and changes in object states.

\begin{figure}
 \centering
 \includegraphics[width=\linewidth]{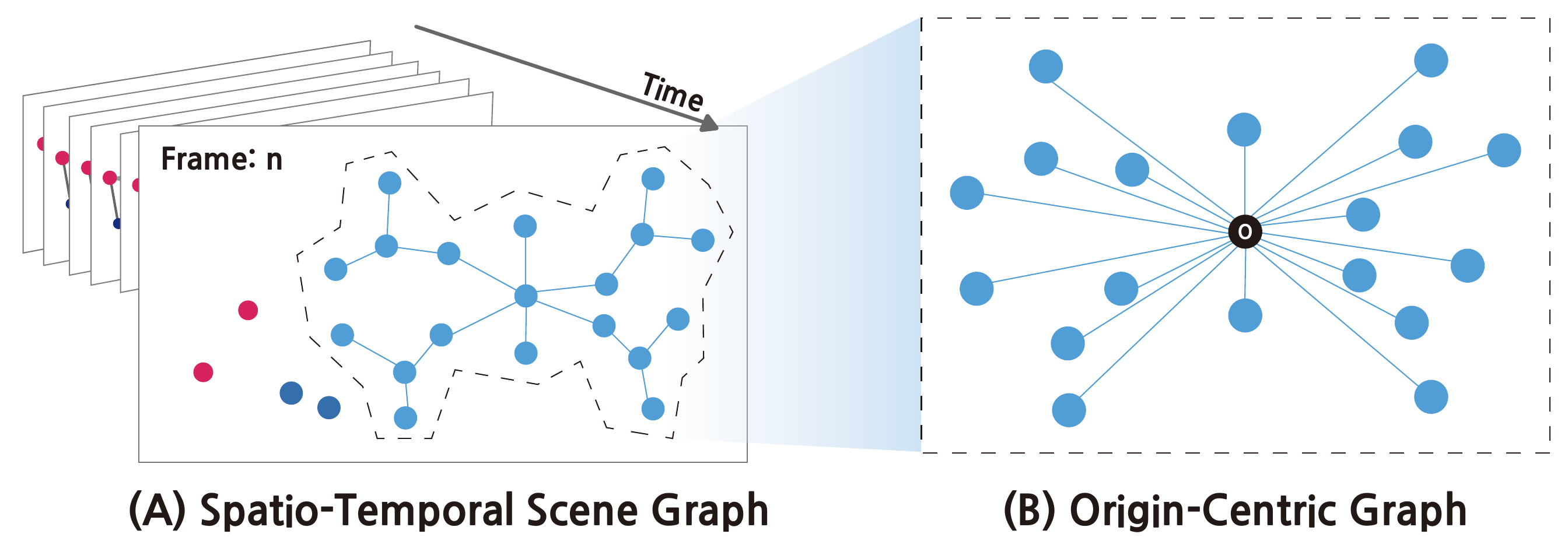}
 \caption{An illustration of constructing an Origin-centric Graph by identifying the origin object from the fully assembled STSG. In (B), “O” denotes the origin object.}
 \label{fig:OCG}
\end{figure}

\begin{figure*}[ht]
 \centering
 \includegraphics[width=\linewidth]{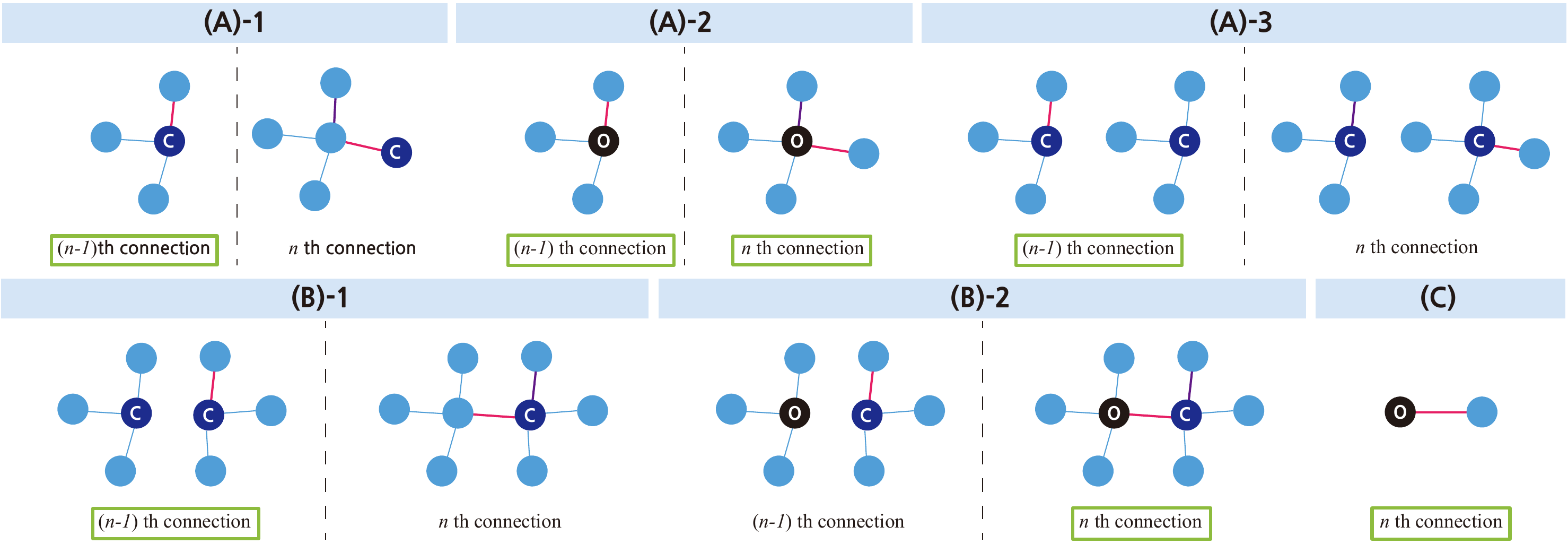}
 \caption{Examples of task breakpoints. (A) Parts connect to the origin object; (B) the central object is updated; (C) a new group begins. Green boxes indicate connections identified as breakpoints.
The dark blue circle labeled “O” represents the origin object, the blue circle labeled “C” indicates the central object, and light blue circles represent general objects.
Pink solid lines show parts assembled at the current moment, while purple solid lines represent parts assembled just before the current moment.}
 \label{fig:BPDetection}
\end{figure*}

\subsection{Origin-Centric Graph}
OCG is a graph designed to represent the structural changes surrounding the central object during the assembly process and to capture the final connectivity relationships among parts in the entire VR recording. As shown in \cref{fig:OCG}, OCG illustrates the relationships between the origin object, which serves as the structural center of the assembly, and other objects. It is constructed offline based on the final assembly stage from the STSG by selecting the origin object using the Adjacency Matrix stored in the STSG and by calculating the shortest paths and degree centrality between the origin object and other nodes. Based on preliminary study results indicating that participants prioritize structural context when setting breakpoints and consider objects closer to the center more important, and supported by previous graph-theoretic analyses emphasizing the structural importance of central objects~\cite{lee2021ikea, slyadnev2020role}, the OCG provides a structured representation of connectivity that reflects both centrality and relational importance. This origin-centered representation characterizes relative structural importance without imposing hierarchical constraints.

Specifically, to select the origin object, we employ degree centrality~\cite{freeman1977set}, denoted as $C_D(o_i)$, which is defined based on the number of immediate neighbors directly connected to each node. Degree centrality is expressed as:
\begin{align}
C_D(o_i) = \frac{\text{deg}(o_i)}{N-1},
\label{equ:centrality}
\end{align}
where $\text{deg}(o_i)$ is the degree of object node $o_i$ and $N$ is the total number of nodes. We designate the node with the maximum degree centrality as the origin object, $o_{\text{origin}}$. Once the origin object is determined, the relative importance of the connections between the origin object and the other objects is quantified using the shortest path distance from the origin object. The weight $W_{\text{origin}}(o_i)$ is defined as follows:
\begin{align}
W_{\text{origin}}(o_i) = \frac{1}{1 + d\big(o_{\text{origin}},\, o_i\big)}.
\label{equ:weight}
\end{align}
where $d(o_{\text{origin}}, o_i)$ represents the shortest path distance from the origin object node to other object node $o_i$ (note that the weight for the origin object is always set to 1). This formulation effectively captures the relative structural importance and hierarchical relationships between the origin object and the other objects during the assembly process. 

\Cref{fig:OCG} illustrates the final OCG, which depicts the connectivity among parts—centered on the origin object—based on the computed weights $W_{\text{origin}}(o_i)$ (see \cref{equ:weight}). In practice, the generation of the OCG follows a systematic graph-analytical process to transform raw connectivity data into this structural importance map. First, the system evaluates the degree centrality (\cref{equ:centrality}) of all nodes at the final assembly stage to identify $o_{\text{origin}}$, which serves as the physical and logical anchor of the entire task. Next, using BFS~\cite{moore1959shortest}, the system traverses the graph starting from $o_{\text{origin}}$ to determine the shortest path distance $d(o_{\text{origin}}, o_i)$ for every node. Finally, these distances are utilized to compute the weights $W_{\text{origin}}(o_i)$ (\cref{equ:weight}), effectively quantifying how essential each part is to the core assembly.
All of these operations have linear time and space complexity $O(N + E)$, ensuring that the OCG can be generated efficiently even for complex assemblies with numerous components. By systematically mapping these graph-theoretic metrics, the STSG-based recording method is capable of capturing the dynamic interactions between users and objects as well as the structural changes occurring during the assembly process. Moreover, the introduction of the OCG allows for an effective representation of the assembly task's connectivity based on the distance from the origin object, thereby facilitating adaptive playback.

\begin{figure*}[ht]
 \centering
 \includegraphics[width=\linewidth]{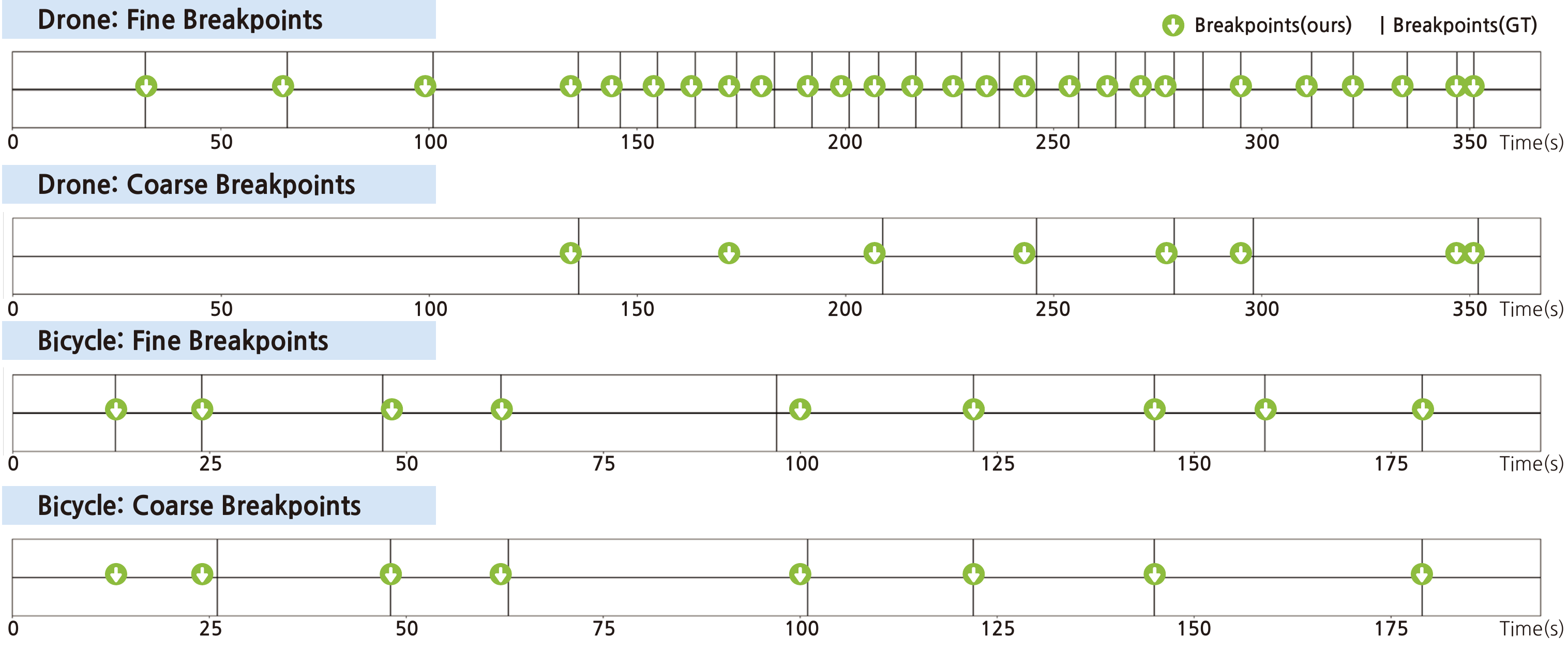}
 \caption{Visualization of the GT breakpoints, derived from user-identified breakpoints in drone and bicycle assembly, alongside the breakpoints identified by our algorithm, for both fine and coarse units.}
 \label{fig:BPResult}
\end{figure*}

\subsection{Task Breakpoints Generation}
We segment the recorded assembly task into fine and coarse units by generating task breakpoints that reflect the structural transition points in the assembly process. A fine breakpoint is defined as a temporal boundary between sequences of frames representing an atomic assembly step that results in a meaningful change in the graph’s topology relative to the origin object. Coarse breakpoints are higher-level boundaries formed by grouping fine units associated with atomic steps that share a common functional goal or central object. This hierarchical segmentation aligns with the mental models of task completion identified in our preliminary study.

\noindent\textbf{Fine Breakpoint Detection}. To identify fine breakpoints, the system monitors the Adjacency Matrix $A_t$ for new assembly events (i.e., when $A_{ij, t-1}=0$ and $A_{ij, t}=1$). A breakpoint is triggered when such a connection satisfies one of the following three structural rules:

\begin{itemize}
    \item \textbf{Direct Connection to Origin (Derived from T1).} A breakpoint is identified when an object $o_i$ or a sub-group connects directly to the origin object $o_{origin}$. As illustrated in \cref{fig:BPDetection}(A), this includes single-to-single, group-to-single, and group-to-group scenarios involving the origin.
    
    \item \textbf{Central Object Update (Derived from T2).} We track the central object ($o_{center}$), defined as the node within the active assembly group with the highest weight $W_{origin}(o_i)$ (see \cref{equ:weight}). If a new connection results in an update of the central object ($o_{center, t} \neq o_{center, t-1}$), it is identified as a structural transition point (\cref{fig:BPDetection}(B)).
    
    \item \textbf{Sub-assembly Formation (Derived from T3).} When a single object $o_i$, previously isolated or connected within a minor group, connects with a new sub-group that does not yet contain the origin, the connection moment is designated as a breakpoint (\cref{fig:BPDetection}(C)).
\end{itemize}

\noindent\textbf{Coarse Breakpoint Detection}. We derive coarse breakpoints, which represent the overall flow of the task, by merging the fine breakpoints identified earlier. 
Based on the preliminary study, we merge consecutive fine units that share the same central object ($o_{center}$) or involve objects of the same category. Specifically, for a sequence of actions involving the same $o_{center}$ (e.g., fastening multiple screws to a single base), the system selects only the final timestamp as the coarse breakpoint. In this manner, the resulting coarse breakpoints effectively consolidate parallel or repetitive tasks into a unified unit, thereby clearly reflecting the overall task flow and its significant transition points. This policy ensures that coarse breakpoints clearly reflect significant transitions in task goals and the broader assembly context, rather than individual physical connections.

\noindent\textbf{Breakpoint Refinement}. In light of the preliminary study findings, which revealed that users perceive the completion of an operator's actions as a meaningful task breakpoint, we refine the initially identified breakpoints to align with the termination of user actions. Since the STSG records the state changes on a per-frame basis, 
we utilize the Hand Adjacency Matrix $H_t$ and user nodes $V_U$ to analyze the continuity of manipulation. Specifically, when a structural connection is detected at frame $t$, the system identifies the final breakpoint $t_{final}$ by searching for the subsequent frame where the user's hands release the involved objects, signified by all relevant entries in $H_t$ returning to zero ($H_{ij} = 0$). This behavior-based refinement process yields breakpoints that more closely match the operator's actual perception rather than the exact moment of physical contact, thereby providing more intuitive task boundaries for adaptive playback.

\section{Evaluation}
\subsection{Study Design and Quantitative Metrics}
For the evaluation, we applied the proposed STSG-based task recording method to two assembly scenarios in VR: bicycle assembly and drone assembly. 
Two expert participants (age: 27 and 30; one female, one male) with prior experience in VR manipulation and interaction were recruited from the KAIST research community. This study was reviewed and approved by the KAIST IRB (No. KH2024-180). A one-hour training session was provided to ensure that the participants were sufficiently familiar with each assembly procedure.
To reflect task complexity in the experiment, we selected the drone assembly as the complex task and the bicycle assembly as the simple task, considering structural factors such as the number of components, repetition, and parallelism~\cite{richardson2006identifying}. The drone scenario involved numerous parts, repeated propeller assembly, and multiple subtasks, resulting in a complex workflow with distributed connections and low centrality of the origin object. In contrast, the bicycle scenario involved fewer components, minimal repetition, and many steps that could be performed in parallel, with most parts directly connected to the origin object (e.g., the bicycle frame), resulting in higher centrality and a clearer hierarchical structure.
Both scenarios were created with reference to real assembly manuals and videos, and for the drone scenario, 3D models based on real physical objects were used to ensure realistic VR representation. The bicycle assembly scenario lasted 3 minutes 4 seconds, and the drone assembly scenario lasted 5 minutes 26 seconds. These two recordings were used as spatial videos in which participants viewed and followed the assembly process, providing data for identifying task segments required for adaptive playback. By evaluating breakpoint detection performance across these representative scenarios, we aimed to confirm the applicability of our method to a variety of assembly tasks.

To establish the GT breakpoints for the two recorded VR task videos, we collected data and annotations from general users. Since the primary goal of breakpoint identification is to support adaptive playback aligned with users’ learning and performance, we adopted user-defined segmentation units as the GT to reflect representative criteria. Following established methodologies in event segmentation research on goal-directed tasks with two levels of granularity (fine and coarse)~\cite{zacks2001perceiving, zacks2001event}, participants carefully reviewed each video and identified breakpoints at moments when a meaningful action ended and the next began. They were instructed to annotate each segment with the corresponding user actions and associated object information.

We then constructed the GT from the collected breakpoint data and annotations. First, breakpoints annotated for the same object manipulation and action were tagged with the same group number, and the data were binned into one-second intervals~\cite{zacks2001perceiving}. To account for slight temporal variations in annotation, DBSCAN clustering~\cite{ester1996density} was applied, constrained to breakpoints with the same group tag to prevent mixing of different events. The clustering parameter 
\(\varepsilon\) was set to 2 seconds, reflecting the temporal ambiguity among human annotators (typically 1–5 seconds) and balancing perceptual characteristics with practical applicability in real scenarios~\cite{shou2021generic}. The minimum number of samples was set to 12, ensuring that only breakpoints agreed upon by a majority of participants were grouped. The median timestamp of each cluster was defined as the GT for the corresponding breakpoint. Through this process, we reliably extracted task unit termination points commonly perceived by multiple users. The final GT dataset was used as the reference standard for evaluating the performance of the proposed breakpoint detection algorithm.

Since traditional tutorial authoring methods rely on manual stepwise segmentation or recording, we set the GT derived from user annotations as the benchmark for evaluation.
To compare the algorithm-detected breakpoints with user-defined GT breakpoints, we employed precision, recall, F1 score, MAE, and RMSE. Precision and recall were calculated by comparing algorithm outputs with GT breakpoints, with the tolerance window determined by the average temporal variance observed among user annotations. Precision measures the proportion of algorithm-detected breakpoints that were correctly matched to GT, while recall measures the proportion of GT breakpoints that were successfully identified by the algorithm. The F1 score, defined as the harmonic mean of precision and recall, captures the overall balance between the two.

MAE and RMSE were computed by pairing each GT breakpoint with the closest predicted breakpoint, with the constraint that a predicted breakpoint could be assigned to only one GT. When a predicted breakpoint overlapped with multiple GT candidates, it was matched to the GT with the smallest temporal error, and unmatched GTs were treated as missed detections. In addition, screenshots of the task units segmented by detected breakpoints were provided to visually illustrate task transitions and segmentation effects. The recording system and scenarios were implemented in Unity 2022.3.21f. The experiments were conducted on a workstation equipped with an AMD Ryzen 9 7950X3D CPU, 64 GB RAM, and an ASUS ROG STRIX RX 4090 GPU, with a Meta Quest~3 used as the VR HMD.

\subsection{Participants and Task Procedures}
A total of 24 new participants with diverse VR experience (age: $M = 25.7$, $SD = 2.7$; 12 females, 12 males), who had not taken part in the preliminary study, were recruited from local undergraduate and graduate students through the official KAIST community website. This study was reviewed and approved by the KAIST IRB (No. KH2024-180).
Participants were instructed to annotate meaningful transition points, referred to as “breakpoints,” in two recorded assembly videos, classifying them into two types (fine and coarse). To balance the experimental conditions, the two assembly videos and the breakpoint types were counterbalanced across participants.
The experiment was conducted in three stages. First, participants wore a Meta Quest~3 headset and watched each VR assembly video from start to finish to familiarize themselves with the overall task flow and structure. Second, they rewatched the same video and annotated fine (or coarse) breakpoints using a timeline slider and the right-hand controller to control playback and pausing. Breakpoints were created by pressing the “+” button at the exact frame when the playback position matched the slider handle, and participants were also provided with a function to delete mistakenly generated breakpoints. Third, coarse breakpoints were annotated in the same way, ensuring that both breakpoint types were created using the same interface. This procedure was repeated for both the drone and bicycle assembly videos, and no participants reported motion sickness during the experiment. Afterward, post-task interviews were conducted to examine participants’ criteria and perceptions of task segmentation. The entire experiment lasted approximately 45 minutes, and each participant received a compensation of \$10.

\begin{table}[ht]
\centering
\caption{Descriptive statistics of user-selected fine and coarse breakpoints across tasks, along with clustered GT breakpoints derived from user annotations.}
\label{tab:breakpoint_stats}
\resizebox{0.85\columnwidth}{!}{%
\begin{tabular}{lcccc}
\toprule
\textbf{Metric} & \multicolumn{1}{c}{\textbf{Drone}} & \multicolumn{1}{c}{\textbf{Drone}} & \multicolumn{1}{c}{\textbf{Bicycle}} & \multicolumn{1}{c}{\textbf{Bicycle}} \\
                & \textbf{Fine} & \textbf{Coarse} & \textbf{Fine} & \textbf{Coarse} \\
\midrule
Mean         & 19.6    & 5.3    & 9.3    & 5.1 \\
Median       & 25      & 5      & 9      & 5 \\
SD           & 10.5    & 2.1    & 2.6    & 1.4 \\
Min          & 5       & 2      & 7      & 3 \\
Max          & 40      & 10     & 18     & 8 \\
Mean BP Range   & 2.81~s    & 6.33~s   & 2.78~s   & 15.14~s \\
\# BP (GT)           & 27      & 6      & 9      & 7 \\
\bottomrule
\end{tabular}%
}
\end{table}

\subsection{Results}
\textbf{User-selected Breakpoints}. ~\cref{tab:breakpoint_stats} summarizes the results of user-identified fine and coarse breakpoints. For each case, Mean, Median, and SD represent the average, median, and standard deviation of the number of breakpoints identified by users. Min and Max denote the minimum and maximum counts, respectively. Mean BP Range indicates the average interval between the earliest and latest breakpoints annotated as the same event, while \#BP (GT) denotes the total number of GT breakpoints obtained by clustering user-identified breakpoints, representing aggregated consensus units.
In the drone scenario, under the fine condition, an average of 19.6 breakpoints was reported ($M=25$, $SD=10.5$, range = 5–40), with \#BP (GT) = 27, and the mean BP range measured 2.81~s. In contrast, the coarse condition produced an average of 5.3 breakpoints ($M=5$, $SD=2.1$, range = 2–10), yielding 6 GT breakpoints, with the mean BP range of 6.33~s.
In the bicycle scenario, under the fine condition, an average of 9.3 breakpoints was reported ($M=9$, $SD=2.6$, range = 7–18), with \#BP (GT) = 9, and the mean BP range of 2.78~s. In contrast, the coarse condition produced an average of 5.1 breakpoints ($M=5$, $SD=1.4$, range = 3–8), with 7 GT breakpoints extracted, and the mean BP range of 15.14~s.

\begin{table}[ht]
\centering
\caption{Performance metrics for breakpoint detection, comparing automatically detected breakpoints (ours) against GT breakpoints.}
\label{tab:bp_performance}
\resizebox{0.9\columnwidth}{!}{%
\begin{tabular}{lcccc}
\toprule
\textbf{Metric} & \textbf{Drone} & \textbf{Drone} & \textbf{Bicycle} & \textbf{Bicycle} \\
                & \textbf{Fine}  & \textbf{Coarse} & \textbf{Fine}    & \textbf{Coarse} \\
\midrule
\# BP (GT)     & 27   & 6    & 9    & 7 \\
\# BP (ours)   & 26   & 8    & 9    & 8 \\
MAE        & 1.38~s & 2.17~s & 0.44~s & 0.57~s \\
RMSE       & 1.66~s & 2.27~s & 1.05~s & 0.93~s \\
\midrule
\multicolumn{5}{l}{\textbf{3-second error range}} \\
Precision      & 0.96 & 1.00 & 1.00 & 1.00 \\
Recall         & 0.96 & 0.75 & 1.00 & 0.88 \\
F1 score       & 0.96 & 0.86 & 1.00 & 0.93 \\
\bottomrule
\end{tabular}%
}
\end{table}

\begin{figure*}[ht]
 \centering
 \includegraphics[width=\textwidth]{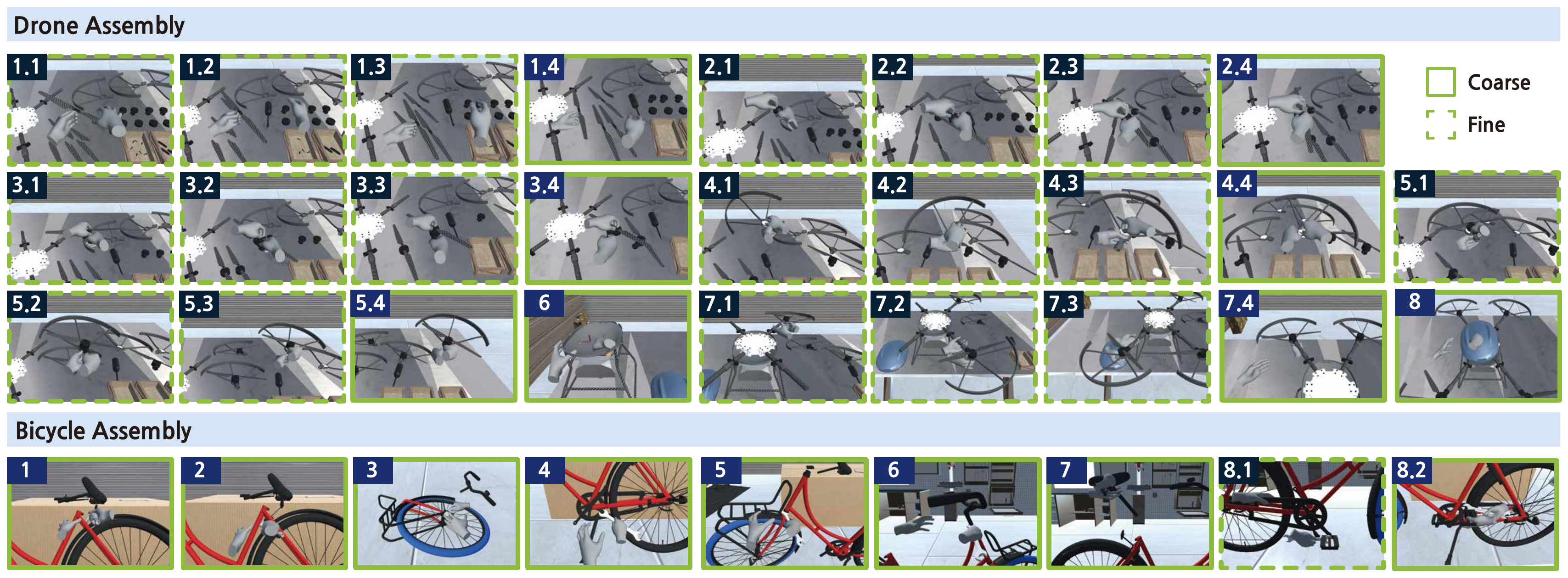}
 \caption{An illustration of the tasks subdivided into fine and coarse units using breakpoints identified by our algorithm. Dashed borders indicate screenshots of specific moments in tasks at the fine-level, while solid borders indicate screenshots of specific moments in tasks at the coarse-level (which also encompass smaller units).}
 \label{fig:Screenshot}
\end{figure*}

\noindent\textbf{Quantitative Metrics}. ~\cref{tab:bp_performance} presents the performance metrics for breakpoint detection under both the Drone and Bicycle scenarios, evaluated under two segmentation conditions (fine and coarse). For each scenario, the table reports the MAE, RMSE, the number of breakpoints based on the GT, and the number of breakpoints detected by ours. 
In the Drone scenario, under the fine condition, an MAE of 1.38~s and an RMSE of 1.66~s were computed, corresponding to 27 GT breakpoints and 26 breakpoints detected by our analysis, while under the coarse condition, an MAE of 2.17~s and an RMSE of 2.27~s were measured, yielding six GT breakpoints with 8 breakpoints detected.
Similarly, in the Bicycle scenario, under the fine condition, an MAE of 0.44~s and an RMSE of 1.05~s were observed, with both GT and our analysis reporting nine breakpoints. 
In contrast, under the coarse condition, an MAE of 0.57~s and an RMSE of 0.93~s were measured, resulting in seven GT breakpoints and 8 breakpoints being detected. 
Furthermore, the evaluation considered a 3~s error range based on the minimum value of the Mean BP Range in~\cref{tab:breakpoint_stats}.
Under this criterion, in the Drone scenario, the coarse condition resulted in precision, recall, and F1 score values of 1.00, 0.75, and 0.86, respectively, while the fine condition produced values of 0.96, 0.96, and 0.96. For the Bicycle scenario the fine condition yielded precision, recall, and F1 score values of 1.00, whereas the coarse condition achieved corresponding values of 1.00, 0.88, and 0.93. 
These results demonstrate that the proposed breakpoint detection algorithm performs competitively within the defined error range, particularly under the fine segmentation condition.

\noindent\textbf{Photo Story.} \cref{fig:Screenshot} shows screenshots of the task segmentation results produced by our algorithm for the drone and bicycle scenarios. In the drone scenario, tasks such as 1.1–1.4 represent the assembly of individual propellers, with fine units capturing detailed actions like attaching propellers and tightening screws, while tasks like 1.4 serve as coarse units that integrate these fine steps. This coarse–fine pattern continues in tasks 2.1–2.4 (motor assembly), 3.1–3.4 (connector attachment), 4.1–4.4 (guard connection), and 5.1–5.4 (propeller module installation), where the final step in each sequence (e.g., 2.4, 3.4, etc.) represents a coarse unit summarizing repeated or grouped fine actions. Tasks 6 and 8, involving frame assembly and top cover installation respectively, are recognized at both fine and coarse units. Similarly, in the bicycle scenario, key steps such as attaching the rear light (task 1), assembling the rear brake (task 2), connecting the wheel (task 3), fixing the stand (task 5), and installing the saddle (task 6) are segmented with fine and coarse granularity. Task 8.1–8.2 covers pedal assembly, with 8.2 functioning as a coarse unit that integrates the fine units.

\section{Discussion}
\subsection{Analysis}
The proposed VR recording method demonstrates its effectiveness in capturing and replaying both user-object relationships and the flow of user actions. The primary goal of our approach is to accurately record hand movements, object interactions, and state changes through a scene graph, thereby supporting task-segmented VR recordings for adaptive playback. The recorded spatial video can be decoded frame by frame and replayed in a VR HMD.
In the evaluation study, most participants reported that experiencing the recorded tasks in VR provided a high level of immersion and facilitated task comprehension, primarily due to the multi-view advantages of spatial video. Participants noted that such recordings could make task execution easier and more intuitive. One participant commented: \textit{``Viewing from a first-person perspective felt like performing the task myself, and being able to see from any 3D angle helped me understand the process without missing details''} (P12). These findings suggest that scene graph-based data recording faithfully reflects user movements and interactions in VR. Consequently, the proposed method effectively supports both the recording and playback aspects of spatial video, delivering the intended functionality for adaptive task guidance.

To effectively record and segment meaningful units in assembly tasks, the proposed OCG played an important role in clearly representing task structures. Building on findings from the preliminary study, where many participants identified breakpoints based on structural changes around a central object and emphasized its importance in the task flow, the OCG was designed to select the task's central object according to degree centrality and to intuitively represent the task structure by using the relative distance between objects. Unlike conventional timeline-based recording methods, this approach provides contextual information on structural interactions and action transitions, thereby managing task complexity more effectively. Furthermore, the combination of OCG and STSG recording explicitly reflects the information that participants consistently evaluated as critical in assembly tasks, contributing to more precise task segmentation and adaptive playback in VR recording.

As shown in \cref{tab:bp_performance}, precision, recall, and F1 scores were generally high across conditions, yet differences emerged depending on task complexity. In more complex tasks, users exhibited diverse and dispersed criteria for segmentation, resulting in greater variability in breakpoint identification. In particular, in the drone scenario, where structural centrality was low and the number of components was large with distributed assembly paths, the algorithm tended to show relatively lower detection performance. By contrast, in the bicycle scenario, the smaller number of components, higher centrality, and simpler structure with more parallelizable steps allowed the algorithm to capture key transitions more reliably. These findings indicate that the proposed OCG effectively captures task hierarchy and centrality, while the STSG explicitly records user-object and object-object interactions. Consequently, the algorithm demonstrated robust segmentation performance that faithfully reflects users’ cognitive structures, even across tasks of varying complexity.

Our breakpoint generation algorithm was found to effectively segment not only fine units but also larger coarse units that capture the overall task flow. In the fine segmentation, the algorithm successfully identified subtle action transitions perceived as meaningful by users, while in the coarse segmentation, it accurately captured broader task structures and major transitions. As illustrated in \cref{fig:Screenshot}, the repetitive small units of propeller assembly were consistently integrated into a larger unit during playback. This indicates that both taskbreakpoint detection and overall structural consistency were maintained in alignment with the GT-based task segmentation criteria. Through this approach, the fine and coarse units segmented by our algorithm naturally correspond to higher-level subtasks and their lower-level steps, which reflect users’ perception of goal-directed tasks. Furthermore, post-experiment interviews revealed that many participants considered coarse units particularly effective for understanding the overall workflow, whereas fine breakpoints were deemed more useful when performing detailed or complex operations. One participant remarked: \textit{``Coarse units are useful for understanding the overall process, while fine breakpoints are more suitable when detailed instructions are required''} (P9). These findings suggest that in an adaptive playback system, coarse breakpoints can provide structural transitions and an overview of the task, while fine breakpoints capture micro-level action shifts, thereby supporting task replay at levels tailored to user proficiency and progress.

The error evaluation using MAE and RMSE confirmed that the breakpoints predicted by our algorithm were highly consistent with the GT breakpoints. Across both the bicycle and drone scenarios, the average MAE and RMSE values remained within the acceptable tolerance range, indicating that the proposed algorithm effectively managed temporal deviations. As shown in \cref{tab:breakpoint_stats}, the Mean BP Range for coarse units averaged 10.7~s, which is substantially broader than the 2.8~s observed for fine units. This explains the slightly higher errors observed in the coarse condition. Overall, these error levels were within an acceptable range and demonstrated that the algorithm’s predictions reliably aligned with the GT. The overarching goal of our VR recording system was to generate spatial videos of VR tasks in a form that supports adaptive playback. Since the segmentation of tasks in VR recordings can vary depending on users’ perception and criteria (for example, recognizing action completion or task-level goals), the exact breakpoint positions may differ depending on the purpose of the recording. Our proposed task breakpoint generation method provides reference points within this perceptual range, enabling task units to be segmented around them. This process significantly reduces the time and cost compared to manual segmentation while still producing VR recordings suitable for adaptive task learning.

\subsection{Limitations}
This study was limited to assembly scenarios for task recording and segmentation for adaptive playback. While the proposed OCG effectively captured object relationships and task progress within assembly workflows, it infers structural importance based on the recorded final assembly topology, which may limit its applicability to unstructured or highly dynamic tasks. In such non-deterministic settings, identifying a single origin object can become ambiguous, potentially requiring extensions such as multi-centric OCGs or alternative graph formulations. More broadly, task-specific factors may differ across other domains. In such cases, additional graph structures or recording strategies may be required, and extending the system to support a broader range of task types remains an important direction for future work.

Furthermore, the proposed algorithm identified breakpoints based on users’ perception of hierarchical task structures at both fine and coarse levels, and the preliminary study confirmed high precision, recall, and F1 scores. However, the evaluation was conducted under assembly scenarios with a limited number of participants and was primarily restricted to comparisons with GT annotations. To comprehensively validate the utility of fine and coarse breakpoints in real adaptive playback scenarios, additional user studies are needed. In particular, evaluations with end users performing tasks with adaptive playback would clarify the impact of hierarchical segmentation on task execution and learning outcomes, such as task completion time and performance improvement.

Finally, this study was conducted in a VR simulation environment and did not utilize real-world spatial video recordings. While the VR environment allowed the performance of the task segmentation algorithm to be evaluated under controlled conditions, extension to augmented reality video recordings is necessary. For automatic task segmentation in real-world work environments, technologies for tracking spatial and object recognition, inter-object connections, and state changes of manipulated objects must be incorporated. The overall framework of scene graph–based recording and task segmentation established through our VR recording method should be retained, but it is necessary to extend the algorithm to AR scenarios where real and virtual content are integrated.

\section{Conclusion and Future Work}

In this study, we proposed a task recording system designed to enable adaptive playback by effectively capturing user-object interactions and task progress, and by segmenting tasks into meaningful units through a breakpoint detection algorithm. To achieve this, we employed a STSG to systematically record user interactions over time and introduced the OCG to clearly represent task structures. Evaluation experiments demonstrated that the approach successfully partitions tasks into hierarchical units, namely fine and coarse units as perceived by users. The algorithm achieved high precision, recall, and F1 scores, indicating strong performance. In particular, the proposed breakpoint-based post-processing step contributed to efficient adaptive playback by reducing both time and cost. These findings provide a solid foundation for developing practical recording and playback systems that can be applied to diverse VR environments and complex tasks.

Although the current method is optimized for assembly tasks, future work should extend the algorithm to a broader range of task domains by incorporating factors such as spatial characteristics and time-sensitive operations. Additional user studies are needed to examine how fine and coarse units support adaptive playback under varying levels of user performance and task complexity, and to assess how this distinction influences user experience and task execution. Finally, since the present study was conducted in a VR simulation environment, future work should leverage AR-based spatial video recording and analysis to capture real-world variability and complexity. Extending the system in these directions will enable more practical, user-centered, and immersive adaptive playback across diverse scenarios.

\acknowledgments{%
	This work was supported by Institute of Information \& communications Technology Planning \& Evaluation (IITP) grant funded by the Korea government (MSIT) (No. RS-2024-00397663, Real-time XR Interface Technology Development for Environmental Adaptation), Institute of Information \& communications Technology Planning \& Evaluation(IITP) grant funded by the Korea government(MSIT) (RS-2019-II191270, WISE AR UI/UX Platform Development for Smartglasses), and the MSIT(Ministry of Science and ICT), Korea, under the Graduate School of Metaverse Convergence support program(IITP-2022(2025)-RS-2022-00156435) supervised by the IITP(Institute for Information \& Communications Technology Planning \& Evaluation).%
}

\bibliographystyle{abbrv-doi-hyperref}

\bibliography{template}

\appendix 

\end{document}